\documentclass[aps, prd, amsmath, floats, floatfix, twocolumn, nofootinbib, superscriptaddress]{revtex4-1}
\usepackage{graphicx}
\usepackage{color}
\usepackage{latexsym}
\usepackage[colorlinks]{hyperref}

\newcommand{\be}{\begin{eqnarray}}
\newcommand{\ee}{\end{eqnarray}}

\begin{document}

\title{A study of the strong gravity region of the black hole in GS~1354--645}

\author{Yerong~Xu}
\affiliation{Center for Field Theory and Particle Physics and Department of Physics, Fudan University, 200438 Shanghai, China}

\author{Sourabh~Nampalliwar}
\affiliation{Theoretical Astrophysics, Eberhard-Karls Universit\"at T\"ubingen, 72076 T\"ubingen, Germany}

\author{Askar~B.~Abdikamalov}
\affiliation{Center for Field Theory and Particle Physics and Department of Physics, Fudan University, 200438 Shanghai, China}

\author{Dimitry~Ayzenberg}
\affiliation{Center for Field Theory and Particle Physics and Department of Physics, Fudan University, 200438 Shanghai, China}

\author{Cosimo~Bambi}
\email[Corresponding author: ]{bambi@fudan.edu.cn}
\affiliation{Center for Field Theory and Particle Physics and Department of Physics, Fudan University, 200438 Shanghai, China}
\affiliation{Theoretical Astrophysics, Eberhard-Karls Universit\"at T\"ubingen, 72076 T\"ubingen, Germany}

\author{Thomas~Dauser}
\affiliation{Remeis Observatory \& ECAP, Universit\"{a}t Erlangen-N\"{u}rnberg, 96049 Bamberg, Germany}

\author{Javier~A.~Garc{\'\i}a}
\affiliation{Cahill Center for Astronomy and Astrophysics, California Institute of Technology, Pasadena, CA 91125, USA}
\affiliation{Remeis Observatory \& ECAP, Universit\"{a}t Erlangen-N\"{u}rnberg, 96049 Bamberg, Germany}

\author{Jiachen~Jiang}
\affiliation{Institute of Astronomy, University of Cambridge, Cambridge CB3 0HA, United Kingdom}

\begin{abstract}
It is thought that the spacetime metric around astrophysical black holes is well described by the Kerr solution of Einstein's gravity. However, a robust observational evidence of the Kerr nature of these objects is still lacking. Here we fit the X-ray spectrum of the stellar-mass black hole in GS~1354--645 with a disk reflection model beyond Einstein's gravity in order test the Kerr black hole hypothesis. We consider the Johannsen metric with the deformation parameters $\alpha_{13}$ and $\alpha_{22}$. The Kerr metric is recovered for $\alpha_{13} = \alpha_{22} = 0$. For $\alpha_{22} = 0$, our measurements of the black hole spin and of the deformation parameter $\alpha_{13}$ are $a_* > 0.975$ and $-0.34 < \alpha_{13} < 0.16$, respectively. For $\alpha_{13} = 0$, we find $a_* > 0.975$ and $-0.09 < \alpha_{22} < 0.42$. All the reported uncertainties are at 99\% of confidence level for two relevant parameters.
\end{abstract}

\maketitle


\section{Introduction}
Einstein's theory of general relativity (GR hereafter) is the standard model for describing effects of gravity in our Universe today. Predictions of GR range from corrections at the scale of Earth~\cite{LLR,GPB} and solar system~\cite{cassini} to entirely new horizons~\cite{pulsar,LIGO}. While the theory has been largely successful in explaining the observations, lingering issues~\cite{dark1,dark2} motivate us to look beyond. Besides observations, there are theoretical reasons to expect a future theory that will supersede GR~\cite{n0,n1,n2,n3,n4}. Thus, determining the domain of validity of GR is important from both observational and theoretical reasons. Many alternative theories have been proposed in this regard. Since GR has been rigorously tested in the weak-field regime, these alternative theories have the same predictions as GR for in this regime and present deviations only when gravity becomes strong.

Black holes are an important prediction of GR. They are expected to be formed in core collapse of massive stars, and as the final product of mergers of massive objects. Other avenues include supermassive black holes, that reside at center of galaxies, and intermediate mass black holes, possibly present in globular clusters. Being extremely compact (i.e., the ratio $M/R$ being close to 1), black holes are the ideal candidates for testing GR in the so-called strong field regime~\cite{r1,r2,r3,r4}.

In 4-dimensional GR, uncharged black holes are completely described by two parameters, which are the ``mass'' $M$ and the ``spin angular momentum'' $J$ of the object. This is the well-known result of the ``no-hair theorem'', which holds under specific assumptions~\cite{h1,h2}. There is also a ``uniqueness theorem'', asserting that the only uncharged black hole solution in 4-dimensional Einstein's gravity is the Kerr metric. One approach then to test GR is to check whether the spacetime metric around astrophysical black holes is indeed described by the Kerr geometry~\cite{test1,test2,test3,test4}.\footnote{This approach has the limitation that if the spacetime is indeed described by the Kerr metric, we cannot differentiate between GR and alternative theories, since most alternative theories include the Kerr metric as a solution~\cite{kerrr1,kerrr2}.}

We have developed a framework recently to look for deviations away from the Kerr metric using X-ray reflection spectroscopy~\cite{noi1,shenzhen}. The astrophysical system consists of a compact object which is described by a parametrically deformed Kerr metric (quantized by \textit{deformation parameters}), with an accretion disk and a corona. In Refs.~\cite{noi1,shenzhen} we introduce the model we developed for use in Xspec~\cite{arnaud} which is the standard software for X-ray data analysis. In Ref.~\cite{noi2}, we presented the first constraints on one of the deformation parameters introduced in~\cite{j-m} using observations of the Narrow Line Seyfert~1 Galaxy 1H0707--495. Constraints using another Narrow Line Seyfert~1 Galaxy, Ark~564, were reported in~\cite{noi3}. And in Ref.~\cite{noi4}, we report constraints using the X-ray binary GX~339--4. The current paper is an analysis in this series. Here we present individual constraints on two parameters of~\cite{j-m} using observations of the low-mass X-ray binary GS~1354--645. We also discuss some issues of the astrophysical modeling that arise naturally in such an analysis.

The outline of the paper is as follows. We review the non-Kerr metric employed in this work and the astrophysical model in Sec.~\ref{sec:review}. Sec.\ref{sec:obs} describes the source, the observation and data reduction. Spectral analysis and results of fitting are reported in Sec.~\ref{sec:models}. Results and some systematic issues are discussed in Sec.~\ref{sec:discuss}. We conclude in Sec.~\ref{sec:conclusion}. Throughout the paper, we adopt the convention $c=G_{\rm N}=1$.

\section{Review\label{sec:review}}
Here we review only the most important aspects. More details on all the aspects can be found in~\cite{noi1,shenzhen}.

The metric we explore is an extension of the Kerr metric, proposed in~\cite{j-m}. We employ such a metric because it has some interesting properties with respect to other parametric black hole spacetimes proposed in the literature. It is everywhere regular outside of the event horizon. It has three independent constants of motion, and thus admits first-order equations of motion as in the Kerr spacetime. As shown in~\cite{j-m}, the metric can reproduce some known black hole solutions in modified theories of gravity (Einstein-Dilaton-Gauss-Bonnet gravity, Chern-Simons gravity, and braneworld models) for specific choices of the values of its deformation parameters.

In Boyer-Lindquist coordinates, the line element of the Johannsen metric reads (we use units in which $G_{\rm N} = c = 1$)~\cite{j-m}
\be\label{eq-jm}
ds^2 &=&-\frac{\tilde{\Sigma}\left(\Delta-a^2A_2^2\sin^2\theta\right)}{B^2}dt^2 \nonumber\\
&&-\frac{2a\left[\left(r^2+a^2\right)A_1A_2-\Delta\right]\tilde{\Sigma}\sin^2\theta}{B^2}dtd\phi \nonumber\\
&&+\frac{\tilde{\Sigma}}{\Delta A_5}dr^2+\tilde{\Sigma}d\theta^2\nonumber\\
&&+\frac{\left[\left(r^2+a^2\right)^2A_1^2-a^2\Delta\sin^2\theta\right]\tilde{\Sigma}\sin^2\theta}{B^2}d\phi^2
\ee
where 
\be
a &=& J/M, \quad \tilde{\Sigma} = \Sigma + f \, \nonumber, \\
\Sigma &=& r^2 + a^2 \cos^2\theta \, , \quad
\Delta = r^2 - 2 M r + a^2 \,\nonumber , \\
B &=& \left(r^2+a^2\right)A_1-a^2A_2\sin^2\theta \, \nonumber, 
\ee
and
\be
f &=& \sum^{\infty}_{n=3}\epsilon_n\frac{M^n}{r^{n-2}} \, , \nonumber\\
A_1 &=& 1 + \sum_{n=3}^{\infty}\alpha_{1n}\left(\frac{M}{r}\right)^n \, , \nonumber\\
A_2 &=& 1 + \sum_{n=2}^{\infty}\alpha_{2n}\left(\frac{M}{r}\right)^n \, , \nonumber\\
A_5 &=& 1 + \sum_{n=2}^{\infty}\alpha_{5n}\left(\frac{M}{r}\right)^n  \, .
\ee
We restrict our attention to the first order deformation parameters $\alpha_{13}$ and $\alpha_{22}$.\footnote{These two parameters have the strongest effect on the iron line~\cite{noi1}, so we restrict our analysis to these parameters.} The Kerr metric is recovered when $\alpha_{13}=\alpha_{22}=0$. In the Kerr metric, black holes only exist for $| a_* | \le 1$, where $a_* = a/M = J/M^2$ is the dimensionless spin parameter. In the Johannsen spacetime, we still have the condition $| a_* | \le 1$ for the existence of the event horizon. In order to exclude a violation of Lorentzian signature in the metric and the existence of closed time-like curves in the exterior region, we have to impose~\citep{j-m}
\be
\label{eq:boundary1}
\alpha_{13} &>& - \left(1 + \sqrt{1 - a_*^2} \right)^3 \, , \\
\alpha_{22} &>& - \left(1 + \sqrt{1 - a_*^2} \right)^2 \, .
\ee
Additionally, the metric is singular for $B = 0$, and we have thus to require that $B$ never vanishes outside of the event horizon~\citep{noi3}. For $\alpha_{22} = 0$, we have the constraint 
\be
\label{eq:boundary2}
\alpha_{13} > - \frac{1}{2} \left( 1 + \sqrt{1 - a^2_*} \right)^4 \, .
\ee
For $\alpha_{13} = 0$, the constraint is
\be
\label{eq:boundary3}
\alpha_{22} < \frac{\left( 1 + \sqrt{1 - a^2_*} \right)^4}{a_*^2} \, .
\ee 
Our exploration of the spin and the deformation parameters happens within these constraints.

The astrophysical system is modeled by a black hole surrounded by a disk and a corona~\cite{corona}. The disk is modeled as a Novikov-Thorne type geometrically-thin optically-thick disk. The corona is described by a broken power-law spectra.

{\sc relxill\_nk}, an extension of {\sc relxill}~\cite{ref1,ref2}, models the X-ray reflection spectrum of black holes described by the Johannsen metric in the disk-corona scenario. It uses the transfer function approach, introduced in~\cite{cunn}, described in~\cite{noi1}. During each analysis, we allow one of the deformation parameters (i.e., one of $\alpha_{13}$ and $\alpha_{22}$) to vary, setting all others at zero.

\section{Observations and data reduction\label{sec:obs}}

GS~1354--645, alias BW~Cir, is a low-mass X-ray binary comprising of a black hole and a low-mass stellar companion. Using optical spectroscopy, the masses are dynamically confirmed at $M_{{\rm BH}} \ge 7.6 \pm 0.7$~$M_\odot$ and $M_{\rm c} \le 1.2$~$M_\odot$~\cite{casares}. The source was discovered in its 1987 outburst by the Japanese X-ray mission \textsl{Ginga}~\cite{makino}. This was followed by another outburst in 1997, detected by \textsl{RXTE}~\cite{brocksopp}. A third outburst was detected in June 2015 by \textsl{Swift}/BAT~\cite{miller}.

During the 2015 outburst, \textsl{NuSTAR} observed GS~1354--645 for three sessions: on June~13 (hereafter Obs 1) 
for about 24~ks, on July~11 (hereafter Obs 2) 
for about 30~ks, and on August~6 (hereafter Obs 3)
for about 35~ks. The first two observations are studied in~\cite{gs}. They found a truncated inner edge for the accretion disk, at $700^{+200}_{-500} R_{{\rm ISCO}}$ for Obs 1, while Obs 2 gave tightly constrained inner edge, close to $R_{{\rm ISCO}}$. No analysis of Obs 3 has been reported yet.

In our analysis, we employed Xspec v12.9.1~\cite{arnaud}. We processed the data from both the FPMA and FPMB instruments using {\it nupipeline} v0.4.5 with the standard filtering criteria and the \textsl{NuSTAR} CALDB version 20171002. We used the {\it nuproducts} routine to extract source spectra, responses, and background spectra. For the source, we chose a circular region of radius 148~arc seconds. For the background, we chose a circular region of radius 148~arc seconds on the same chip. All spectra were binned to a minimum of 30~counts before analysis to ensure the validity of the $\chi^2$ fit statistics. In what follows, we report the analysis of Obs 2 only. For Obs 1, we find that the inner edge of the accretion disk is truncated at large radii, in agreement with what found in~\cite{gs} and making it unsuitable to constrain the background metric.

\section{Spectral analysis\label{sec:models}}
We fit Obs 2 with the following models:

{\bf Model~1:} {\sc tbabs*powerlaw}.

\begin{figure}[h]
\vspace{0.5cm}
\begin{center}
\includegraphics[type=pdf,ext=.pdf,read=.pdf,width=0.49\textwidth]{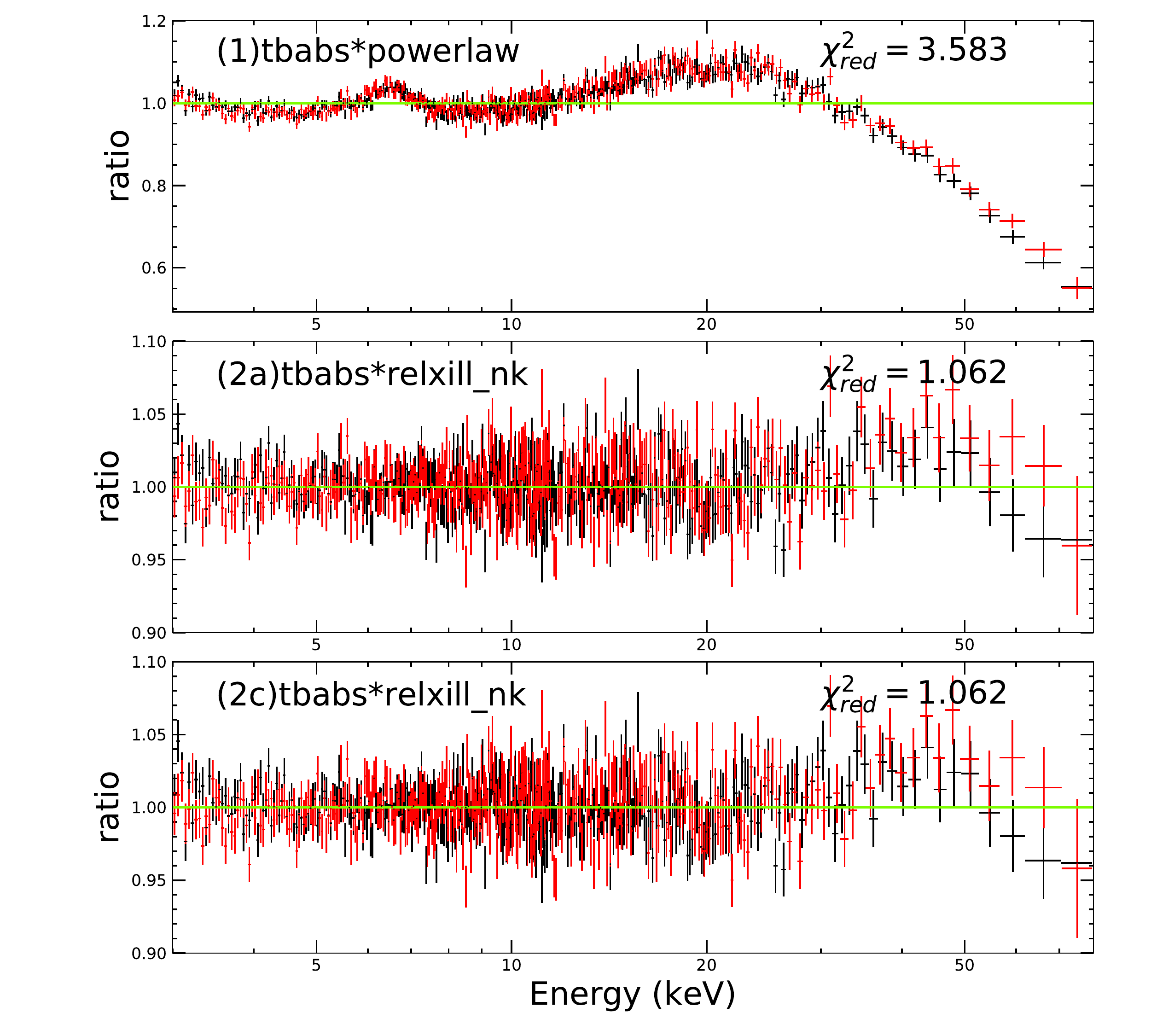}
\end{center}
\vspace{-0.5cm}
\caption{Data to best-fit model ratio for model~1 (top panel), model~2a (middle panel) and model~2c (bottom panel). Black crosses are used for FPMA and red crosses are used for FPMB. See text for more details. \label{fig:1a2a_ratio}}
\end{figure}

This is the simplest model. {\sc tbabs} describes the Galactic absorption~\citep{wilms} and we fix the galactic column density to $N_{\rm H} = 0.7 \cdot 10^{22}$~cm$^{-2}$, which we obtain from the HEASARC column density tool, based on~\cite{dickey90}. {\sc powerlaw} describes the power-law spectrum from the hot corona around the black hole. The best-fit values are reported in the second column in Tab.~\ref{t-fit}, where we see that the minimum of the reduced $\chi^2$ is around~3.6. Fig.~\ref{fig:1a2a_ratio} (top panel) shows the data to the best-fit model ratio, and we clearly see a broad iron line at 6-7~keV and a Compton hump at 10-30~keV.

\begin{table*}
\centering
\vspace{0.5cm}
\begin{tabular}{lccccc}
\hline\hline
Model & 1 & 2a & 2b & 2c & 3 \\
\hline
{\sc tbabs} &&&&& \\
$N_{\rm H} / 10^{22}$ cm$^{-2}$ & \hspace{0.6cm} 0.7$^\star$ \hspace{0.6cm} & \hspace{0.6cm} 0.7$^\star$ \hspace{0.6cm} & \hspace{0.6cm} 0.7$^\star$ \hspace{0.6cm} & \hspace{0.6cm} 0.7$^\star$ \hspace{0.6cm} & \hspace{0.6cm} 0.7$^\star$ \hspace{0.6cm} \\
\hline
{\sc powerlaw} &&&&& \\
$\Gamma$ & $1.541^{+0.001}_{-0.001}$ & -- & -- & -- & -- \\
\hline
{\sc relxill\_nk} &&&&& \\
$q_{\rm in}$ & -- & $10.0_{-2.1}$ & $9.97_{-0.09}$ & $10.0_{-1.5}$ & $10.0_{-0.8}$ \\
$q_{\rm out}$ & -- & $0.5^{+0.9}$ & $0.48_{-0.30}^{+0.06}$ & $3^\star$ & $0.7_{-0.6}^{+0.3}$ \\
$R_{\rm br}$ [$M$] & -- & $4.4_{-0.7}^{+3.1}$ & $4.39_{-0.04}^{+3.13}$ & $1.95^{+0.12}_{-0.10}$ & $3.9^{+0.7}_{-0.7}$\\
$i$ [deg] & -- & $77.7^{+8.9}_{-1.1}$ & $77.7_{-0.3}^{+1.0}$ & $74.6^{+0.9}_{-1.6}$ & $78.0^{+0.5}_{-1.1}$  \\
$a_*$ & -- & $> 0.992$ & $> 0.992$ & $> 0.995$ & $0.995_{-0.005}$\\
$R_{\rm in}$ [$R_{\rm ISCO}$] & -- & 1$^\star$ & $1.000^{+0.013}$ & 1$^\star$ & 1$^\star$  \\
$\alpha_{13}$ & -- & $0.0565^{+0.053}_{-0.307}$ & $0.0565^{+0.014}_{-0.297}$ & $-0.2935^{+0.003}_{-0.036}$ & --  \\
$\alpha_{22}$ & -- & -- & -- & -- & $-0.0077^{+0.235}_{-0.075}$ \\
$\Gamma$ & -- & $1.657^{+0.007}_{-0.013}$ & $1.657^{+0.011}_{-0.009}$ & $1.654^{+0.012}_{-0.012}$ & $1.657^{+0.012}_{-0.005}$ \\
$\log\xi$ & -- & $2.40_{-0.04}^{+0.06}$ & $2.40_{-0.04}^{+0.06}$ & $2.43_{-0.06}^{+0.10}$ & $2.40_{-0.04}^{+0.06}$  \\
$A_{\rm Fe}$ & -- & $0.52^{+0.06}$ & $0.52^{+0.06}$ & $0.59^{+0.06}_{-0.06}$ & $0.52^{+0.06}$  \\
$E_{\rm cut}$ [keV] & -- & $162_{-14}^{+11}$ & $162_{-13}^{+11}$ & $166^{+17}_{-8}$ & $162^{+17}_{-13}$ \\
$R$ & -- & $1.27_{-0.08}^{+0.09}$ & $1.27_{-0.08}^{+0.12}$ & $1.05_{-0.08}^{+0.08}$ & $1.31^{+0.11}_{-0.11}$ \\
\hline
$\chi^2$/dof & 9782.07/2730 & 2887.78/2720 & 2887.78/2719 & 2890.14/2721 & 2887.86/2720\\
& = 3.58318 & = 1.06168 & = 1.06207 & = 1.06216 & = 1.06171 \\
\hline\hline
\end{tabular}
\vspace{0.2cm}
\caption{Summary of the best-fit values for model~1 (simple power-law), model~2a (power-law with reflection component and $\alpha_{13}$ free), model~2b (as model~2a but with $R_{\rm in}$ free), model~2c (as model~2a but with $q_{\rm out} = 3$ frozen) and model~3 (as model~2a but with the deformation parameters $\alpha_{13} = 0$ and $\alpha_{22}$ free). The reported uncertainties correspond to the 90\% confidence level for one relevant parameter. $^\star$ indicates that the parameter is frozen. See the text for more details. \label{t-fit}}
\end{table*}

{\bf Model~2a:} {\sc tbabs*relxill\_nk} with $\alpha_{13}$ as the deformation parameter.

\begin{figure*}[t]
\begin{center}
\vspace{0.3cm}
\includegraphics[type=pdf,ext=.pdf,read=.pdf,width=0.47\textwidth]{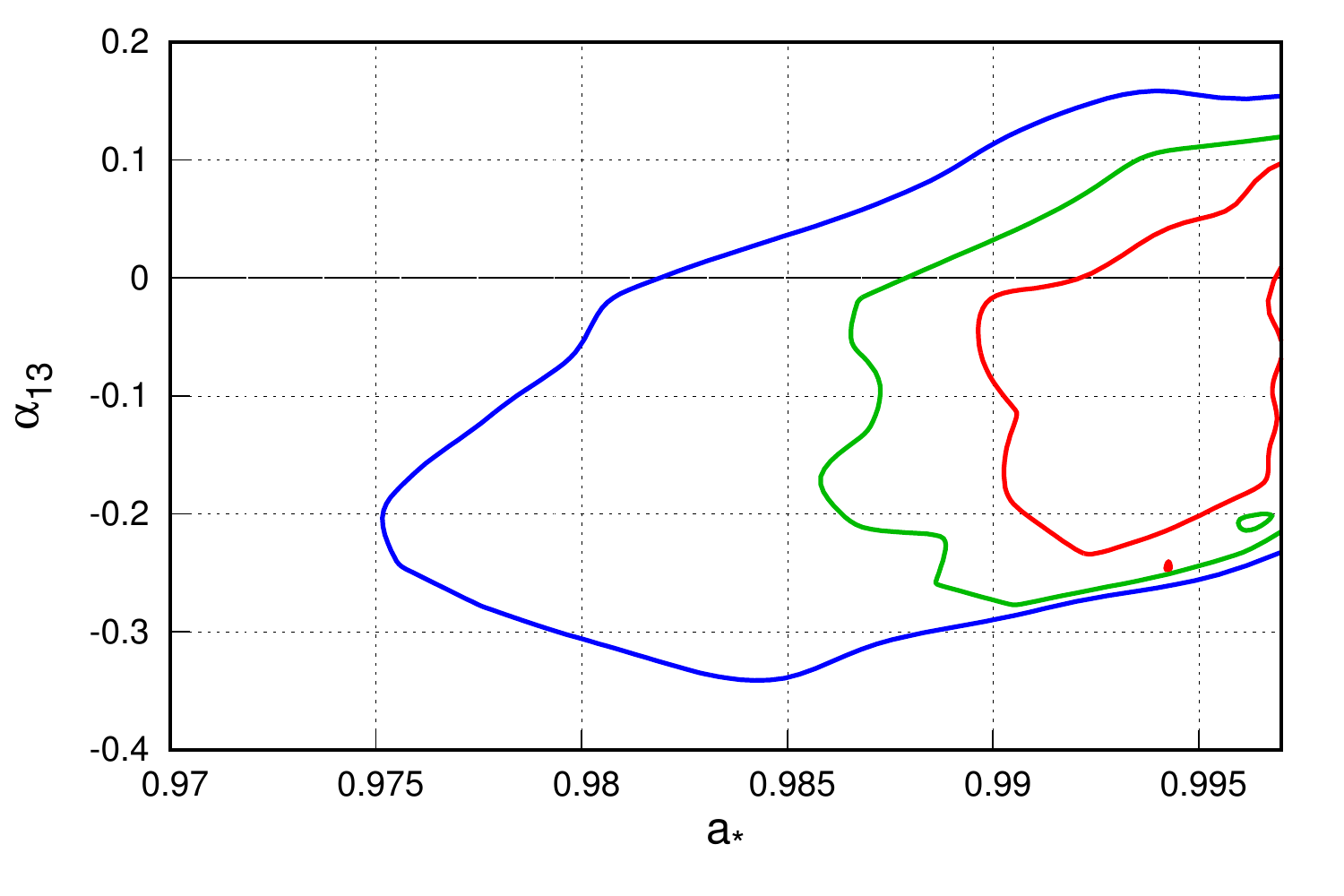}
\hspace{0.5cm}
\includegraphics[type=pdf,ext=.pdf,read=.pdf,width=0.47\textwidth]{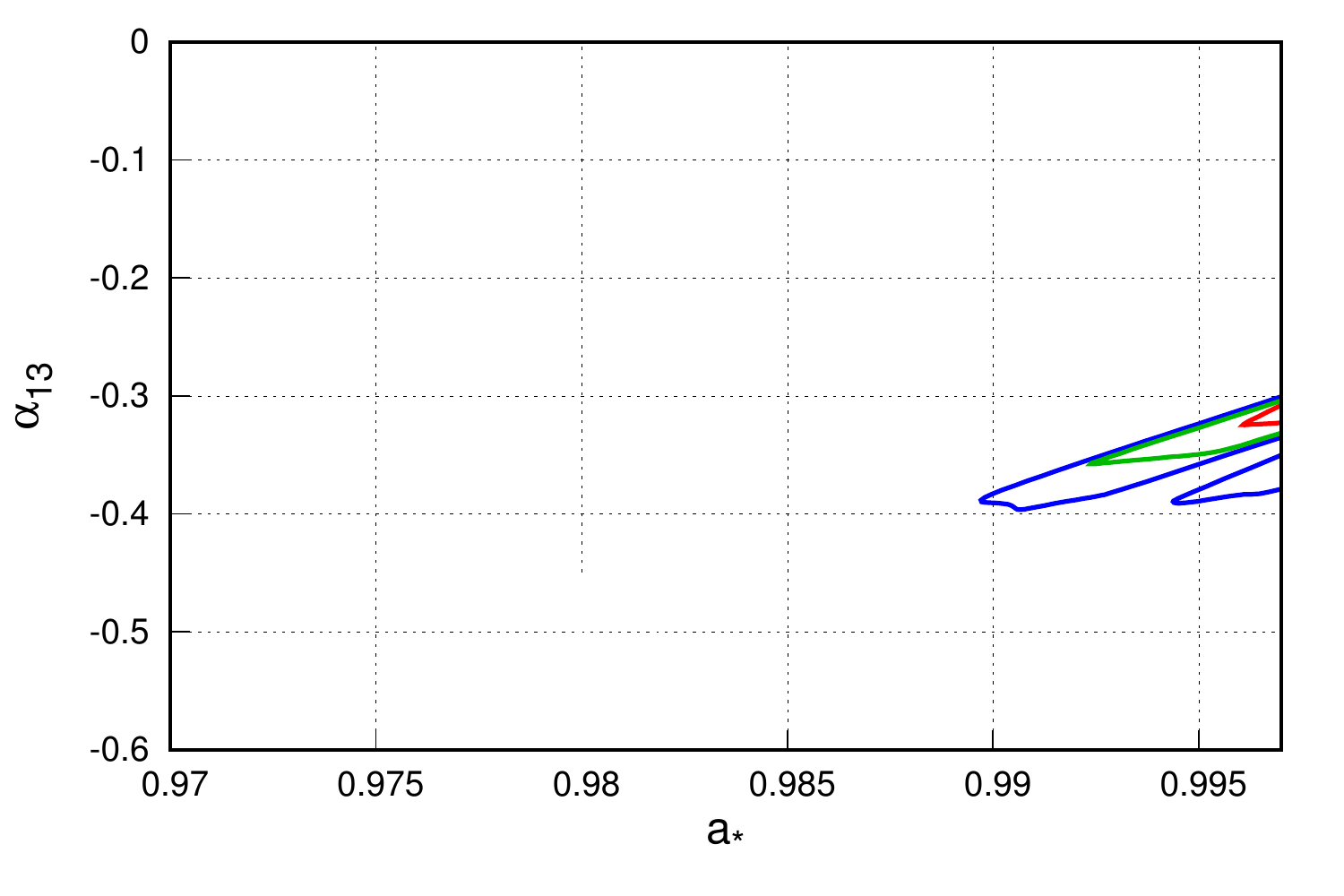}
\end{center}
\vspace{-0.6cm}
\caption{Constraints on the spin parameter $a_*$ and on the Johannsen deformation parameter $\alpha_{13}$ from model~2a (left panel) and model~2c (right panel). The red, green, and blue lines indicate, respectively, the 68\%, 90\%, and 99\% confidence level contours for two relevant parameters. The solid black line marks the Kerr solution.\label{f-2a2c}}
\end{figure*}

We now use our model {\sc relxill\_nk}. The emissivity profile of the disk is modeled with a broken power-law and is described by three free parameters: the inner emissivity index $q_{\rm in}$, the outer emissivity index $q_{\rm out}$, and the breaking radius $R_{\rm br}$. The inner edge of the accretion disk, $R_{\rm in}$, is set at the innermost stable circular orbit (ISCO) of the metric. Tab.~\ref{t-fit} shows the best-fit values along with statistical uncertainties at the 90\% confidence level. Fig.~\ref{fig:1a2a_ratio} shows the data to the best-fit model ratio (middle panel). We know that the spin parameter and the deformation parameter have some degeneracy; we report this degeneracy between $a_*$ and $\alpha_{13}$ using a contour plot in Fig.~\ref{f-2a2c} (left panel).

{\bf Model~2b:} same as model~2a but with the location of the inner edge, $R_{\rm in}$, free. 

While it is a standard assumption to place the inner edge of the disk at the ISCO, in this model we leave $R_{\rm in}$ free and fit for it. The best-fit values are reported in the fourth column in Tab.~\ref{t-fit}. In particular, we find that $R_{\rm in}$ prefers a value very close to $R_{\rm ISCO}$. This validates the assumption and for other models, we always set $R_{\rm in} = R_{\rm ISCO}$.

{\bf Model~2c:} same as model~2a but with $q_{\rm out} = 3$. 

We impose $q_{\rm out} = 3$, which is the value expected in the lamppost coronal geometry\footnote{This is a common model for the corona, see Ref.~\cite{ref1} for details.} at large radii without light bending. The best-fit values are reported in the fifth column in Tab.~\ref{t-fit} and the ratio of data to the best-fit model is shown in Fig.~\ref{fig:1a2a_ratio} (bottom panel). The right panel in Fig.~\ref{f-2a2c} shows the degeneracy contours of $a_*$ vs $\alpha_{13}$. As we can see, the Kerr metric is not recovered in this case. This shows the importance of the emissivity profile in testing the Kerr metric. We discuss this point in more detail in Sec.~\ref{sec:discuss}.

{\bf Model~3:} {\sc tbabs*relxill\_nk} with $\alpha_{22}$ as the deformation parameter. 

\begin{figure}[t]
\begin{center}
\vspace{0.3cm}
\includegraphics[type=pdf,ext=.pdf,read=.pdf,width=0.47\textwidth]{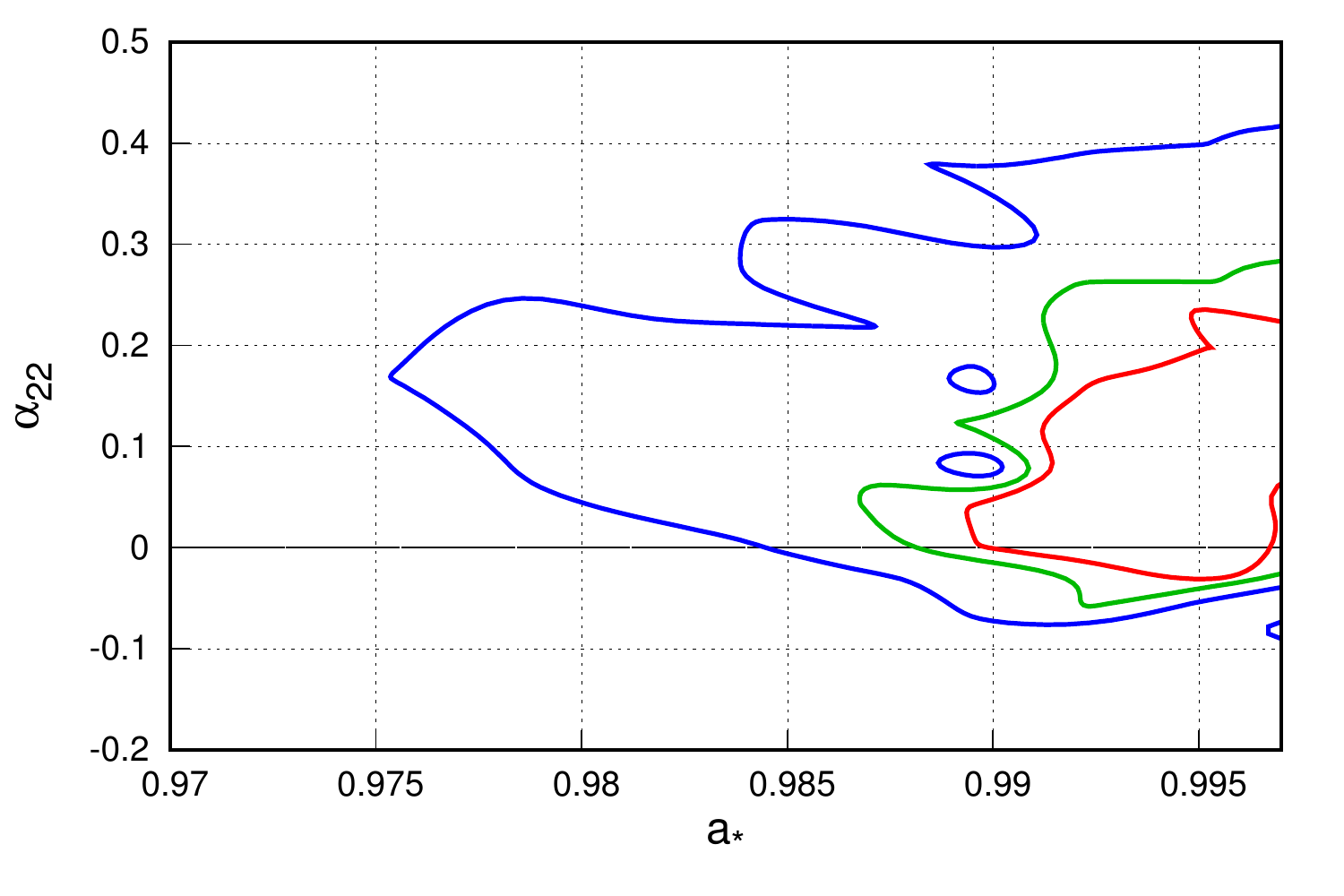}
\end{center}
\vspace{-0.6cm}
\caption{Constraints on the spin parameter $a_*$ and on the Johannsen deformation parameter $\alpha_{22}$ from model~3. The red, green, and blue lines indicate, respectively, the 68\%, 90\%, and 99\% confidence level contours for two relevant parameters. The solid black line marks the Kerr solution. \label{f-alpha22epsilon3}}
\end{figure}

With this model we constrain the deformation parameter $\alpha_{22}$. The assumptions are the same as in model~2a, with $R_{\rm in} = R_{\rm ISCO}$ and $q_{\rm out}$ free. The best-fit values are reported in the sixth column in Tab.~\ref{t-fit}. Fig.~\ref{f-alpha22epsilon3} shows the degeneracy contours of $a_*$ vs $\alpha_{22}$.

\section{Discussion\label{sec:discuss}}
We now discuss some aspects of the analysis presented above in particular and of the technique in general. To begin, our best-fit model (model~2a) for $\alpha_{13}$ provides the following constraints:
\be\label{eq-model2a}
a_* > 0.975 \, , \quad
-0.34 < \alpha_{13} < 0.16 \, ,
\ee
at $99\%$ confidence level. These constraints on $\alpha_{13}$ are the strongest one obtained until now with our reflection model {\sc relxill\_nk}, see~\cite{noi2,noi3,noi4,shenzhen}, and it is consistent with the Kerr metric. The strength of the present constraint can to be attributed to the following factors. Firstly, the source analyzed here is a black hole binary observed with \textsl{NuSTAR}, whereas in Refs.~\cite{noi2,noi3} the sources are supermassive black holes and therefore the observations there have a lower photon count. Secondly, in~\cite{noi4}, although the source is an X-ray binary, the spin of the hole is not as high as reported here and the inner edge of the disk (assumed to be co-located with the ISCO) in~\cite{noi4} does not extend as close to the horizon as it does in the present case. The nature of the ISCO contours in the $a_*-\alpha_{13}$ phase space (see~\cite{j-m,noi1}) implies that the degeneracy between $a_*$ and $\alpha_{13}$ decreases as $a_*$ increases. Therefore, with a high spin estimate in the present case, we obtain stronger constraints on $\alpha_{13}$.

Our constraints on the second deformation parameter, $\alpha_{22}$, come from model~3:
\be
a_* > 0.975 \, , \quad
-0.09 < \alpha_{22} < 0.42 \, ,
\ee
at $99\%$ confidence level. These constraints on $\alpha_{22}$ are also stronger than previous constraints~\cite{noi3}. Apart from the fact that the source in~\cite{noi3} is a supermassive black hole, the best-fit spin is lower and the location of the ISCO is farther in that paper than reported here. Again, the nature of the ISCO contours in $a_*-\alpha_{22}$ phase space (see~\cite{j-m}) imply that the degeneracy between $a_*$ and $\alpha_{22}$ decreases as $a_*$ increases. Therefore, a higher spin estimate in the present case provides better constraints on $\alpha_{22}$.

All the above results have accounted for the statistical uncertainty. Systematic uncertainties on the other hand are difficult to estimate and there is currently no detailed study in the literature to help to quantify them in the measurements employing reflection spectroscopy. We illustrate one aspect of systematic uncertainty with model~2c. In this model, we fix the outer emissivity index, $q_{\textrm{out}}$, at 3 to mimic a specific coronal geometry, the lamppost. In the lamppost model, the corona is a point source located on the black hole spin axis at a few gravitational radii (more sophisticated geometries allow for an extended lamppost, a tube along the spin axis). For such a geometry, the emissivity index in the outer part of the accretion disk is equal to 3. We report the best-fit parameters for model~2c in Tab.~\ref{t-fit} and the data to model ratio plot in Fig.~\ref{fig:1a2a_ratio} (bottom panel). Nothing remarkable is apparent here: the reduced $\chi^2$ value for this model is comparable to other models, and the plot shows no obvious residuals (cf. the middle panel in Fig.~\ref{fig:1a2a_ratio}). The only outlier is the constraint on $\alpha_{13}$. Fig.~\ref{f-2a2c} in the right panel shows the confidence regions in the $a_*-\alpha_{13}$ phase space. The Kerr solution ($\alpha_{13}=0$) is strongly violated. In the present case, since we artificially fix the parameter $q_{\textrm{out}} = 3$, we can be confident that the exclusion of the Kerr solution is a systematic effect. Analysis with a version of \textsc{relxill\_nk} that incorporates the lamppost geometry will be better to judge whether such an exclusion indeed happens with the lamppost geometry with this source. Development of the lamppost version of \textsc{relxill\_nk} is currently underway and we hope to resolve this conundrum in near future.

A second source of systematic uncertainty could be the assumption of a Novikov-Thorne type thin disk in \textsc{relxill\_nk}. This is thought to be correct with a good approximation for sources accreting between the 5\% and the 30\% of their Eddington limit~\cite{penna,ssp}. In the case of GS~1354--645, the distance is poorly constrained and should be between 25 and 61~kpc~\cite{casares}. For its mass, we have the lower bound $M \ge 7.6 \pm 0.7$~$M_\odot$~\cite{casares}. The resulting luminosity is $L/L_{\rm Edd} \le 0.53$, which is consistent with the 5-30\% range but it also allows higher and lower luminosities. For higher luminosities, the gas pressure may not be negligible and the inner edge of the disk may be at a radius smaller than the ISCO one, leading to an overestimate of the black hole spin.

\section{Concluding remarks\label{sec:conclusion}}

In a series of papers on precision tests of the Kerr solution with X-ray reflection spectroscopy, we have developed a model for X-ray data analysis, namely \textsc{relxill\_nk}, that incorporates parametrically deformed Kerr metrics in a disk-corona setup, and we have applied this model to several X-ray observations to obtain constraints on the deformation parameters which mark departure from the Kerr solution. In the current study, we have applied \textsc{relxill\_nk} (introduced in~\cite{noi1}) to the X-ray binary GS~1354--645. We report constraints on the deformation parameters $\alpha_{13}$ and  $\alpha_{22}$ introduced in~\cite{j-m}. Constraints obtained here are stronger than those reported previously, in~\cite{noi2,noi3,noi4}. We have discussed some aspects of the constraints on deformation parameters and some systematic issues associated with this technique.

Precision tests of the Kerr geometry with X-ray reflection spectroscopy is a nascent field and significant scope for improvement exists. With new X-ray satellites planned for launch in near future~\cite{extp}, we expect data of much higher quality. Improvements in \textsc{relxill\_nk}, including other disk geometries, coronal geometries and non-Kerr metrics, are underway.


{\bf Acknowledgments --}
This work was supported by the National Natural Science Foundation of China (NSFC), Grant No.~U1531117, and Fudan University, Grant No.~IDH1512060. S.N. acknowledges support from the Excellence Initiative at Eberhard-Karls Universit\"at T\"ubingen. A.B.A. also acknowledges the support from the Shanghai Government Scholarship (SGS). C.B. and J.A.G. also acknowledge support from the Alexander von Humboldt Foundation. J.J. is supported by the Cambridge Trust and the Chinese Scholarship Council Joint Scholarship Programme (201604100032).


\end{document}